\input harvmac.tex

\lref\malda{J.~ M.~Maldacena, ``The Large N Limit of Superconformal
Field Theories and Supergravity", hep-th/9711200.}%
\lref\witholo{E.~Witten, ``Anti De Sitter Space And Holography",
hep-th/9802150.}
\lref\witthe{E.~Witten, ``Anti-de Sitter Space, Thermal Phase Transition,
And Confinement In Gauge Theories'', hep-th/9803131.}
\lref\w{E.~Witten, ``Baryons And Branes In Anti de Sitter Space",
hep-th/9805112.}%
\lref\ks{S.~Kachru, E.~Silverstein, ``4d Conformal Field Theories
and Strings on Orbifolds", hep-th/9802183.}
\lref\lnv{A.~Lawrence, N.~ Nekrasov, C.~ Vafa,
``On Conformal Theories in Four Dimensions", hep-th/9803015.}%
\lref\dgmr{M.~R.~Douglas, G.~Moore, ``D-branes, Quivers, and ALE
Instantons", hep-th/9603167.}
\lref\dhvw{L.~Dixon, J.~A.~Harvey, C.~Vafa and E.~Witten,
\np B 261 (1985) 678; \np B 274 (1986) 285.}
\lref\dgm{M.~R.~Douglas, B.~ R.~Greene, D.~ R.~Morrison, ``Orbifold
Resolution by D-Branes", Nucl. Phys. {\bf B506} (1997) 84.}
\lref\grw{S.~Gukov, M.~Rangamani and E.~Witten,
``Dibaryons, Branes, And Strings In AdS Orbifold Models", hep-th/9811048.}
\lref\nshorizon{D.~R.~Morrison, M.~R.~Plesser, ``Non-Spherical
Horizons, I", hep-th/9810201.}
\lref\kutasov{A.~Giveon, D.~Kutasov, ``Brane Dynamics and Gauge
Theory", hep-th/9802067.}

\lref\polchinski{J.~Polchinski, ``Dirichlet Branes And Ramond-Ramond
Charges", Phys. Rev. Lett. {\bf 75} (1995) 4724.}
\lref\hmoore{J.~A.~Harvey, G.~Moore, ``On the algebras of BPS
states", Commun. Math. Phys. {\bf 197} (1998) 489.}
\lref\mmoore{R.~Minasian, G.~Moore, ``K-theory and Ramond-Ramond
charge", J.High Energy Phys. {\bf 9711} (1997) 002.}
\lref\eyin{Y.-K.~E.~Cheung, Z.~Yin, ``Anomalies, Branes, and
Currents", Nucl.Phys. {\bf B517} (1998) 69.}
\lref\senzero{A.~Sen, ``Stable Non-BPS States In String Theory",
JHEP {\bf 6} (1998) 7.}
\lref\senone{A.~Sen, ``Stable Non-BPS Bound States Of BPS
$D$-branes", hep-th/9805019.}
\lref\sentwo{A.~Sen, ``Tachyon Condensation On The Brane Antibrane
System", hep-th/9805170.}
\lref\bergman{O.~Bergman and M.~R.~Gaberdiel, ``Stable Non-BPS
$D$-particles", hep-th/9806155.}
\lref\senthree{A.~Sen, ``$SO(32)$ Spinors Of Type I And Other Solitons
On Brane-Antibrane Pair", hep-th/9808141.}
\lref\senfour{A.~Sen, ``Type I D-particle and its Interactions",
hep-th/9809111.}
\lref\wk{E.~Witten, ``D-Branes And K-Theory", hep-th/9810188.}
\lref\wbound{E.~Witten, ``Bound States Of Strings And
$p$-Branes", Nucl. Phys. {\bf B460} (1996) 335.}
\lref\sensethi{A.~Sen, S.~Sethi, ``The Mirror Transform of Type I
Vacua in Six Dimensions", Nucl. Phys. {\bf B499} (1997) 45.}
\lref\atiyah{M.~F.~Atiyah, ``K-Theory", W. A. Benjamin,
New York, 1967.}
\lref\abs{M.~F.~Atiyah, R.~Bott and A.~Shapiro, ``Clifford
Modules", Topology {\bf 3} (1964) 3.}
\lref\atiyahr{M.~F.~Atiyah, ``K-Theory And Reality", Quart. J. Math.
Oxford (2) {\bf 17} (1966) 367.}
\lref\ah{M.~F.~Atiyah and F.~Hirzebruch, ``Vector Bundles
and Homogeneous Spaces", Proc. of Symposia in Pure Math.,
Differential Geometry, Amer. Math. Soc. 1961, 7.}
\lref\korp{M.~Fujii, ``$K_O$-Groups of Projective Spaces",
Osaka J. Math. {\bf 4} (1967) 141.}
\lref\klens{T.~Kambe, ``The structure of $K_{\Lambda}$-rings
of the lens space and their applications",
J. Math. Soc. Japan {\bf 18} (1966) 135.}
\lref\kslens{T.~Kobayashi, M.~Sugawara, ``$K_{\Lambda}$-Rings
of Lens Spaces $L^n (4)$", Hiroshima Math. J. {\bf 1} (1971) 253.}
\lref\ksplens{T.~Kawaguchi, M.~Sugawara, ``$K-$ and $KO$-Rings
of the Lens Space $L^n(p^2)$ for Odd Prime $p$",
Hiroshima Math. J. {\bf 1} (1971) 273.}
\lref\ksym{H.~Minami, ``$K$-Groups of Symmetric Spaces I, II",
Osaka J. Math. {\bf 12} (1975) 623, Osaka J. Math. {\bf 13}
(1976) 271.}
\lref\tor{R.~Morelli, ``K Theory of a Toric Variety",
Adv. in Math. {\bf 100} (1993) 154.}
\lref\sharpe{A.~Knutson, E.~Sharpe, ``Sheaves on Toric
Varieties for Physics", hep-th/9711036.}
\lref\adams{J.~F.~Adams, ``Vector Fields on Spheres",
Ann. of Math., {\bf 75} (1962) 603.}
\lref\eqlect{``Equivariant K-Theory", Lectures by M.~F.~Atiyah and
G.~B.~Segal, Coventry: University of Warwick.}
\lref\segal{G.~B.~Segal, ``Equivariant K-Theory", Inst. Hautes Etudes
Sci. Publ. Math. No.34 (1968) 129.}
\lref\atseg{M.~F.~Atiyah and G.~B.~Segal, ``Equivariant
K-Theory and Completion", J. Diff. Geom. {\bf 3} (1969) 1.}
\lref\gp{E.~ G.~ Gimon, J.~ Polchinski, ``Consistency Conditions
for Orientifolds and D-Manifolds", Phys. Rev. {\bf D54} (1996)
1667, hep-th/9601038.}
\lref\cliff{I.~R.~Porteous, ``Clifford Algebras and the Classical
Groups", Cambridge Univ. Press.}
\lref\karone{M.~Karoubi, ``K-theory. An introduction",
Springer-Verlag, Berlin-New York, 1978.}
\lref\kartwo{M.~Karoubi, ``Algebres de Clifford et K-theorie",
(French) Ann. Sci. Ecole Norm. Sup. (4) {\bf 1} (1968) 161.}
\lref\askew{M.~F.~Atiyah and I.~M.~Singer, ``Index Theory
for Skew-Adjoint Fredholm Operators", Inst. Hautes Etudes
Sci. Publ. Math. No.37 (1969) 305.}

\lref\newsen{A.~Sen, ``BPS D-branes on Non-supersymmetric
Cycles", hep-th/9812031.}
\lref\newhorava{P.~Horava, ``Type IIA D-Branes, K-Theory, and
Matrix Theory", hep-th/9812135.}
\lref\howu{Pei-Ming Ho, Yong-Shi Wu, ``Brane Creation in
M(atrix) Theory", Phys. Lett. {\bf B420} (1998) 43.}
\lref\holiwu{P.~-M.~Ho, M.~Li, Y.~-S.~Wu, ``P-P' Strings
in M(atrix) Theory", Nucl. Phys. {\bf B525} (1998) 146.}
\lref\bdfive{M.~Berkooz, M.~R.~Douglas, ``Five-branes in
M(atrix) Theory", Phys. Lett. {\bf B395} (1997) 196.}
\lref\tasi{J.~Polchinski, ``TASI Lectures on D-Branes",
hep-th/9611050.}
\lref\bgs{C.~ P.~Bachas, M.~ B.~Green, A.~Schwimmer,
``(8,0) Quantum mechanics and symmetry enhancement in type I'
superstrings", J.High Energy Phys. {\bf 9801} (1998) 006.}
\lref\gglcone{M.~B.~Green, M.~Gutperle, ``Light-cone
supersymmetry and D-branes", Nucl. Phys. {\bf B476} (1996) 484.}
\lref\berggl{O.~Bergman, M.~R.~Gaberdiel, G.~Lifschytz,
``Branes, Orientifolds and the Creation of Elementary
Strings", Nucl.Phys. {\bf B509} (1998) 194.}
\lref\atanomaly{M.~F.~Atiyah, ``Eigenvalues of the Dirac
Operator", Lecture Notes in Math. 1111, Springer-Verlag (1985) 251.}
\lref\indexone{M.~F.~Atiyah and I.~M.~Singer, ``The index of
elliptic operators: I", Ann. of Math. {\bf 87} (1986) 484.}
\lref\indextwo{M.~F.~Atiyah and G.~B.~Segal, ``The index of
elliptic operators: II", Ann. of Math. {\bf 87} (1986) 531.}
\lref\berry{``Geometric Phases in Physics", edited by
F.~Wilczek and A.~Shapere, World Scientific, 1989.}

\font\cmss=cmss10 \font\cmsss=cmss10 at 7pt

\def\IB{\relax\hbox{$\inbar\kern-.3em{\rm B}$}}
\def\IC{\relax\hbox{$\inbar\kern-.3em{\rm C}$}}
\def\ID{\relax\hbox{$\inbar\kern-.3em{\rm D}$}}
\def\IE{\relax\hbox{$\inbar\kern-.3em{\rm E}$}}
\def\IF{\relax\hbox{$\inbar\kern-.3em{\rm F}$}}
\def\IG{\relax\hbox{$\inbar\kern-.3em{\rm G}$}}
\def\IGa{\relax\hbox{${\rm I}\kern-.18em\Gamma$}}
\def\IH{\relax{\rm I\kern-.18em H}}
\def\IK{\relax{\rm I\kern-.18em K}}
\def\IL{\relax{\rm I\kern-.18em L}}
\def\IP{\relax{\rm I\kern-.18em P}}
\def\IR{\relax{\rm I\kern-.18em R}}
\def\Z{\relax\ifmmode\mathchoice
{\hbox{\cmss Z\kern-.4em Z}}{\hbox{\cmss Z\kern-.4em Z}}
{\lower.9pt\hbox{\cmsss Z\kern-.4em Z}}
{\lower1.2pt\hbox{\cmsss Z\kern-.4em Z}}\else{\cmss Z\kern-.4em
Z}\fi}

\def\II{\relax{\rm I\kern-.18em I}}

\def\S{{\bf S}}
\def\B{{\bf B}}
\def\R{{\bf R}}
\def\RP{{\bf RP}}

\def\pt{{\rm pt}}
\def\hf{{1\over 2}}


\def\Cl {{\cal C}l}
\def\CD {{\cal D}}
\def\CE {{\cal E}}

\def\CI {{\cal I}}

\def\CM {{\cal M}}
\def\CN {{\cal N}}
\def\CO {{\cal O}}


\def\p{\partial}

\def\D{{\slash\!\!\!\! D}}



\def\coker{{\mathop {\rm coker}}}

\def\Tr{{\rm Tr}}

\def\p{\partial}

\def\Det{{\rm Det}}

\def\inbar{\,\vrule height1.5ex width.4pt depth0pt}

\def\a{\alpha}

\def\g{\gamma}

\def\m{\mu}
\def\n{\nu}
\def\la{\lambda}

\def\p{\partial}


\def\np{{Nucl. Phys. }}


\hbox{PUPT-1830}
\hbox{ITEP-TH-1/99}
  \Title{ \vbox{\baselineskip12pt\hbox{hep-th/9901042}}}
{\vbox{ \centerline
{K-Theory, Reality, and Orientifolds}
\vskip2pt     
    }}
\centerline{Sergei Gukov\foot{On leave from the Institute of
Theoretical and Experimental Physics and the L.~D.~Landau
Institute for Theoretical Physics}}
\vskip 2pt
\centerline{Joseph Henry Laboratories, Princeton University}
\centerline{Princeton, New Jersey 08544}
\centerline{gukov@pupgg.princeton.edu}
\vskip 50pt

{\bf \centerline{Abstract}}

We use equivariant $K$-theory to classify charges of
new (possibly non-supersymmetric) states localized on
various orientifolds in Type II string theory.
We also comment on the stringy construction of new D-branes
and demonstrate the discrete electric-magnetic duality in
Type I brane systems with $p+q=7$, as proposed by Witten.

\Date{January 1999}

\newsec{Introduction}

During the past few years, D-branes have been
playing a significant role
in the study of non-perturbative dynamics of supersymmetric
string and field theories. Dirichlet $p$-branes are themselves
`solitonic' BPS states charged under Ramond-Ramond fields
\polchinski. In turn, SUSY gauge theories appear as low-energy
description of parallel D-branes \wbound.
In numerous applications (extended) supersymmetry was an
indispensable ingredient to guarantee stability of the vacuum
and to retain control in the strong coupling regime -- for a
review see \kutasov.

The study of non-supersymmetric string vacua is especially important
for making a contact with reality. Some progress in this
direction has been achieved by Sen \refs{\senzero, \senone}~
who found new states in non-BPS brane systems with
tachyon condensation \sentwo.
At least perturbatively, these states are stable because of
charge conservation. For example, Type I D-particle,
dual to the $SO(32)$ heterotic spinor, is the lightest state
with $SO(32)$ spinor quantum numbers \refs{\senthree, \senfour}.
In fact, there are topological obstructions preventing a decay
of such states.

In the systematic approach via $K$-theory \wk, Witten proposed
a new way of looking at D-brane charges \foot{Possible
interpretation of BPS charges in terms of $K$-theory was
first considered in \mmoore.}. The basic idea that
D-branes are equipped with gauge bundles naturally leads to
the identification of lower-dimensional branes with topological
defects (vortices) in the gauge bundle.
Because this argument is purely topological,
it does not rely on supersymmetry at all.
For this reason, it not only reproduces conventional spectrum
of BPS D-branes, but it also yields charges of new non-supersymmetric
states. For example, a novel non-BPS eight-brane, a seven-brane and
a gauge instanton with $\Z_2$ charges were found in Type I string
theory \wk.

In the present paper we classify charges of new (possibly
non-supersymmetric) states in Type II orientifolds by means
of equivariant $K$-theory.
The reason to consider orientifolds rather than orbifolds is
that in many cases $K$-theory of an orbifold does not provide
more information than the ordinary cohomology theory of its
smooth resolution (we present some arguments and examples
in section 6). Thus, in reduction to lower dimension,
D-brane charges follow by wrapping over all possible cycles.
The statement
obviously does not hold if the singularity is `frozen',
{\it i.e.} if it can not be blown up. Such orbifolds correspond to
the non-zero flux of the Neveu-Schwarz anti-symmetric
tensor field \sensethi, which we always assume to vanish.

The paper is organized in such a way that the balance between
physics and mathematics shifts gradually from one section to
another. The next section is a warm-up where we briefly review
the results of \wk, and prepare to study $K$-theory of
orientifolds. Then, we study in details three types
of orientifolds, as in \wk. Section 3 is devoted to
$\Omega$ orientifolds (type $(i)$). Depending on the
choice of projection, D-brane charges
localized on such orientifolds take values in the real
$K$-theory $KR(X)$, or its symplectic analog which we call
$KH(X)$. Calculating these groups we find a number of
new D-branes with $\Z_2$ charges, {\it e.g.} a non-BPS
3-brane localized on an $\CO^{+}5$ plane. Of particular
interest is Type I string theory which is an example
of such orientifolds where the involution acts trivially
in the space-time.
It was proposed in \wk, that there is a $(-1)$ monodromy
experienced by a gauge instanton crossing an 8-brane, or
by a 0-brane winding around a 7-brane. In section 4 we
justify this conjecture in two different ways.
First, we observe Berry's phase analyzing
degeneracy of the 0-7 fermion spectrum.
Second, a gauge-theoretic approach leads to the
spectral flow of the Dirac operator. In section 5
we return to the main theme of the paper and classify
D-brane charges localized on
$(-1)^{F_L}$ orientifolds. The spectrum turns out to
be the same for any dimension of an orientifold.
Hence, the analogy with Type IIA theory can be used to
deduce physical properties of the new states.
Even though in this paper we will not try to present a complete
analysis of $(-1)^{F_L} \cdot \Omega$ orientifolds,
this case will be mentioned in section 6, where some orbifold
models will be discussed as well.
Finally, we will present our conclusions in section 7.

Close to the completion of this work we received
preprints \refs{\newsen, \newhorava} which complement
and slightly overlap discussion of $(-1)^{F_L}$ orientifolds
in section 5, in particular Type IIA string theory.


\newsec{General Aspects}

\subsec{K-theory and D-brane charges}

Before we proceed to the $K$-theory of orientifolds
it is necessary to set notation and formulate
the problem. Consider Type IIB
superstring\foot{Generalization to Type IIA
theory is straightforward, and we comment on that in the
end of each section. In later sections we also clarify
the relation between D-branes in IIA and IIB theories,
regarding the former as $(-1)^{F_L}$ orientifold of the
latter.} propagating in the space-time:
$$
\R^{d+1} \times X
$$
with $n$ 9-branes and $m$ $\bar 9$-branes, the simplest
setup to define $K$-theory of D-brane charges \wk.
For a moment we forget about tadpole cancellation
condition, and impose it later. The nine-branes
are supplemented with gauge bundles $E$ and $F$ respectively.
In order to describe a $d$-brane, we want the configuration
$(E,F)$ to be translationally invariant in $(d+1)$
directions. In other words, $(E,F)$ labels a pair of
bundles over $X$.

Of course, brane -- anti-brane system described above
is unstable which is marked by the presence of a
tachyon $T$ in the
spectrum of open $9$ -- ${\bar 9}$ strings. The tachyon
is a map:
\eqn\tmap{T \colon F \to E}
or put differently, a section of $E \otimes F^{\ast}$.
Therefore, such system tends to annihilate itself
unless there is some topological obstruction.
The latter is measured by the $K$-theory group $K(X)$
which we are about to define.

Assuming that an arbitrary number of brane -- anti-brane pairs
can be created (or annihilated) from vacuum
with isomorphic bundles $H$ and $H'$, we come
to the equivalence relation:
\eqn\erel{ (E,F) \sim (E \oplus H, F \oplus H') }
which makes a semigroup of pairs $(E,F)$ an abelian group
$K(X)$ called Grothendieck group \refs{\atiyah, \karone}.
The additive structure of $K(X)$ is induced by the direct
sum of bundles.

To keep the discussion less abstract, it is instructive
to work out a simple example that will prove useful below.
Let us calculate the Grothendieck group of a point $K(\pt)$.
Any bundle over a point is isomorphic to the trivial bundle of
certain dimension $n$. In this case, the equivalence
\erel\ takes the form: $(n,m) \sim (n+k, m+k)$ where
$n$, $m$ and $k$ are non-negative integers representing
the dimensions of bundles. Therefore, the elements of $K(X)$
are $(n,m) = n-m$ which constitute a group of
integer numbers $\Z$.

Now, using the result $K(\pt) = \Z$, we make a few refinements
of the construction. First of all, we notice that a map of
$X$ to a point induces the homomorphism $\rho \colon
\Z \to K(X)$. Since in physical applications the difference
$(n-m)$ is fixed by the tadpole cancellation condition
we should be actually interested in the cokernel of $\rho$,
the so-called reduced $K$-theory group $\tilde K (X) \equiv \coker~ \rho$.
We also expect a $d$-brane to have finite tension.
This condition translates to the statement that the charges
of the physical D-branes take values in the $K$-theory
with compact support \wk. In other words, it tells that
\tmap\ is an isomorphism outside a set $U \subset X$ such
that the closure $\overline U$ is compact. Physically,
$U$ represents the region in the transverse space where
the $d$-brane is localized.
Since this condition automatically implies
reduced $K$-theory, in the rest of the paper
(except in section 5) we will omit tilde and use the
notation $K(X)$ for the {\it reduced} $K$-theory with
compact support
\foot{There is a nice definition of such $K$-theory
given by G.~B.~Segal in terms of {\it complexes} \segal.
A complex is given by a sequence:
$$
\CE^{k} \colon 0 \to E^0 \to E^1 \to \ldots \to E^{k-1} \to 0
$$
of vector bundles $\{ E^i \}$ over $X$ which fails to be exact
over the compact support $U \subset X$. The complex $\CE$
is called {\it acyclic} if $U=\emptyset$. Then, $K(X)$ is defined
as the set of isomorphism classes of complexes $\CE$ on
$X$ modulo acyclic complexes. Even though it may sound too
abstract, this definition has a clear physical interpretation.
For example, an acyclic complex of length 2 represents a pair
of isomorphic bundles $E \cong F$. Equivalence modulo such
complexes is nothing but the equivalence relation \erel\
which allows the creation/annihilation of brane--anti-brane
pairs with isomorphic bundles. Therefore, at length 2,
we just recover the standard definition of $K(X)$ given in
the text. It might seem that equivalence modulo acyclic
complexes of arbitrary length is stronger than the relation
\erel. However it is not the case \refs{\segal, \eqlect},
and the two definitions are equivalent.
As a next step, acyclic complex $\CE^3$ is given
by the exact sequence: $0 \to E \to G \to F \to 0$. This is
nothing but the charge conservation condition for scattering
of (anti-)BPS states $[E] + [F] \to [G]$ found in terms of the
ordinary cohomology \hmoore.}.
If the space-time is
flat, $X = \R^{9-d}$, then $K(\R^{9-d})$ with compact
support is isomorphic to $K(\S^{9-d})$ by adding a
point `at infinity'. This group is equal to $\Z$ for
$d$-odd, and is trivial otherwise (see {\it e.g.}
\refs{\atiyah, \karone}). Thus, we obtain the standard
spectrum of D-brane charges in Type IIB string theory.
When $d$ is odd, we take $S_{\pm}$ to be positive
(negative) spinor representation of $SO(9-d)$, the
group of rotations in the transverse directions.
Then, the explicit form of the tachyon field
corresponding to the unit $d$-brane charge placed at
the origin of $\R^{9-d}$ can be written in terms of Gamma
matrices $\vec \Gamma \colon S_{-} \to S_{+}$ \wk:
\eqn\tachyon{T(\vec x) = \vec \Gamma \cdot \vec x}
where we omit a suitable normalization factor.

Generalization of this construction to other string
theories is also possible \wk. Here we state without
proof that in Type IIA string theory D-brane charges
take values in $K(X \times \S^1)$, while the charges
of Type I D-branes are measured by $KO(X)$. For details
we refer the reader to the original work \wk. On the other
hand, the necessary mathematical background on $K$-theory
can be found in \refs{\atiyah, \karone, \eqlect}.


\subsec{Equivariant K-theory and Orientifolds}

In what follows we consider space-time of the form:
$$
\R^{p+1} \times (\CM^{9-p} / G)
$$
where $\CM^{9-p}$ is a smooth manifold, and the discrete symmetry
group $G$ acts continuously on $\CM^{9-p}$. Being interested
in the $d$-brane charges, we also consider vector bundles
$E$ over $X = \R^{p-d} \times \CM^{9-p}$,
such that the projection
$E \to X$ commutes with the action of $G$. The above
conditions define the category of $G$-equivariant
bundles over $G$-space $X$ \refs{\segal, \eqlect}. The
corresponding $K$-theory is called $G$-equivariant
$K$-theory $K_G (X)$. In many ways, $K_G (X)$
is similar to the ordinary $K$-theory. For example,
such properties of $K(X)$ like Thom isomorphism and
Bott periodicity continue to hold in the equivariant
case \refs{\segal, \eqlect, \atseg}. Another basic
theorem of equivariant $K$-theory tells that if
$G$ acts freely on $X$, then:
\eqn\freeact{K_G (X) \cong K(X/G)}
This isomorphism will prove to be useful in calculations.

So far we have described $K$-theory of orbifolds \wk.
However, it turns out that, compared to the usual
cohomology theory, for `regular' orbifolds it does not provide
new states ({\it cf.} section 6). For this reason we consider
$G$ accompanied by a world-sheet symmetry action.
We refer to its fixed point set (a number of $\R^{p+1}$ planes)
as orientifold $p$-planes, or $\CO p$-planes for short.

Following the approach of \wk, we address the following question:
What are the charges of stable (possibly non-BPS) states localized
at a singularity of $\CM^{9-p} / G$ ({\it i.e.}
located on the $\CO p$-plane)?
To answer this question, we have to recast it explicitly in
terms of vector bundles --- the language used throughout the paper.
Stability of a $d$-dimensional object just means that it is a
topological defect in the gauge bundle of 9-brane -- antibrane system,
{\it i.e.} its charge takes values in the $G$-equivariant $K$-theory of
$X=\R^{p-d} \times \CM^{9-p}$ \wk. Provided that $d<p$, a $d$-brane
can be constructed from $p$-branes placed at the fixed point of
$\CM / G$. The $d$-brane is stable if it is the lightest state charged
under $p$-brane gauge group \senzero. The condition for such a
state to be localized at a singularity translates to the assertion
that $K$-theory a has compact support which includes the singular
point. Therefore, it has to be $G$-equivariant $K$-theory.
Indeed, if in the vicinity of the singularity the tachyon
is homotopic to the vacuum\foot{In other words, $T \colon F \to E$
is an isomorphism.},
and this region is path-connected to the infinity, then one
can deform the compact support (the core of a gauge `vortex')
arbitrarily far from the singularity.
Therefore, the state is {\it not} localized at the singularity and
is represented by an element of $K_G (X) \cong K(X_G)$ where $G$
acts freely on $X$ \refs{\segal, \atseg}.
Since for the most of our applications this group is isomorphic to
the usual $K$-theory $K(X)$, we consider only the states localized
on the singularity.

Suppose $\CM$ is a vector space, and  $G$
acts on $\CM$ with at most one isolated singularity
at the origin. If we define $\S$ to be a unit sphere
in $\CM$, then the smooth manifold $H = \S / G$
(= unit sphere in $X= \CM / G$) is automatically
Einstein. In analogy with the AdS/CFT correspondence \malda,
it is natural to call it a `horizon', {\it cf.} \nshorizon.
According to \malda, gauge theory
on $p$-branes placed at the singularity is dual to the
supergravity compactification on $H$. The counterpart of
this relation in the equivariant
$K$-theory with compact support is
given by the exact triangle for the pair $(\CM, \S)$:
\eqn\tri{
\matrix{
K^{\ast} (H) & &
\buildrel \delta^{\ast} \over \longrightarrow & &
K^{\ast}_G (\CM, \S) \cr
& \nwarrow && \swarrow & \cr
&&K^{\ast}_G (\CM) &&}}
where $\delta \colon \CM \to \S$ is boundary homomorphism.
To write the equivariant group $K_G (\S)$ we used the
fact that $G$ acts freely on $\S$ and the theorem \freeact.
Because the relative $K$-theory $K^{\ast}_G (\CM, \S)$ is
canonically isomorphic to the $K$-theory with compact
support, the exact sequence \tri\ will prove to
be useful in computations of the groups $K_G (X)$.
In mathematical terminology, $X$ is a cone on $H$,
and $\Sigma' H = X / H$ is called unreduced suspension
of $H$ \refs{\atiyah, \karone}.


\newsec{Orientifolds of type (i):
$\R^{p+1} \times (\CM^{9-p} / \Omega \cdot \CI_{9-p})$}

\subsec{$\tau^2 = -1$: The Real K-theory}

Now we are ready to consider the first example: orientifolds
\eqn\omor{
\R^{p+1} \times (\CM^{9-p} / \Omega \cdot \CI_{9-p}) }
of type $(i)$, as in \wk. In this case the
generator of $G=\Z_2$ is a combination of the involution
$\CI_{9-p}$ on $\CM^{9-p}$
and the world-sheet parity transformation $\Omega$.
String orientation reversal induces an anti-linear
map $\tau \colon E_x \to E_{\tau(x)}$ on the gauge bundle.
There are two consistent orientifold projections in Type IIB
string theory \gp, corresponding to $\tau^2=1$ and $\tau^2=-1$
respectively. In the first case we obtain $KR$-theory \wk,
while in the second case D-brane charge takes values in the
group which we denote\foot{This is in analogy
with symplectic bundles, where $\tau$ is multiplication
by $j$ over the field of quaternions $\IH = \IC \oplus
j \IC$.} by $KH(X)$ and study in the next part of this section.
There are two types of orientifolds, called $\CO^{\pm}$
according to their tadpole contribution. They carry
$\mp 2^{p-4}$ units of $p$-brane charge and produce $SO$
or $Sp$ gauge groups respectively. In what follows, we will
see that the choice of projection is determined by $\tau$
(whether its square is equal to plus or minus identity), so that the
states on the orientifolds are classified by $KR(X)$ or
$KH(X)$.

Let us first consider the case $\tau^2=1$ corresponding to
the quantization of 9-branes with $SO$ Chan-Paton factors.
Our major example in this paper will be the simplest case
$\CM^{9-p} = \R^{9-p}$ where new D-brane charges can be found.
Then, orientifolds \omor\ take the following form:
\eqn\omorr{ \R^{p+1} \times (\R^{9-p} / \Omega \cdot \CI_{9-p}) }
It is convenient to introduce the notation $\R^{p,q}$ for
the space-time $X=\R^q \times \R^p$ with the involution $\CI_p$
acting on the second factor. The convention is chosen
to agree with the notation of the corresponding
linear space in \atiyahr. We also denote:
\eqn\defs{\matrix{
\B^{p,q} \equiv {\rm unit\ ball\ in~} \R^{p,q} \cr
\S^{p,q} \equiv {\rm unit\ sphere\ in~} \R^{p,q} }}
Note, $\S^{p,q}$ has dimension $p+q-1$, {\it e.g.} $\S^{o,n} =
\S^{n-1}$.

In mathematical terms, the above properties define
the real category of vector bundles over $X$ with compact
support. Therefore the $d$-brane charge localized on the
orientifold $p$-plane takes values in the real $K$-theory
\wk, which we denote as:
\eqn\krrel{ KR^{9-p, p-d} (\pt) \equiv
KR(\B^{9-p, p-d}, \S^{9-p, p-d})}
These are the so-called $(p, q)$ suspension groups of a point
\atiyahr; compare with the ordinary definition
$KR^{-n} (X,Y) \equiv KR (X \times \B^{0,n}, X \times \S^{0,n}
\cup Y \times \B^{0,n}) \cong KR (\Sigma^{n} (X/Y))$
\refs{\atiyah, \karone}.
Because the involution acts trivially on a single
point, we find helpful the following general relation:
\eqn\krtoko{KR (X_R) \cong KO(X_R) }
where $X_R$ is the set of fixed points under the
involution $\tau$ \atiyahr.

To calculate \krrel, we also need the following periodicity
isomorphisms established by Atiyah:
\eqn\krisom{ KR (X) \cong KR^{-8} (X) }
$$
KR^{p,q} (X) \cong KR^{p+1,q+1} (X) \cong KR^{p-q} (X)
$$
The first property follows from multiplication by the
generator of $KR^{-8} (\pt)$, while multiplication
by the generator of $KR^{1,1} (\pt)$ induces the second
isomorphism in \krisom. In the special case (of
our interest) when $X=\pt$, one can independently prove the
formulas \krisom\ via the periodicity of
the corresponding Clifford algebras, {\it cf.} section 6.

To compute $KR(\R^{9-p, p-d})$, we use the periodicity theorem
\krisom\ which leads to the group $KR(\R^{0, 2p-d-1})$ of the
real space
$\R^{0, 2p-d-1}$ with a compact support where the involution
acts trivially, $\tau (x) = x$. Hence, by the formula
\krtoko, we obtain for the $d$-brane charges:
\eqn\krfin{KR(\R^{9-p, p-d}) \cong KO(\S^{2p-d-1}) }
Modulo the Bott periodicity, we list all the $KO$-groups of
spheres in the table below \adams:
\vskip 5 pt
$$
\def\tbntry#1{\vbox to 23 pt{\vfill \hbox{#1}\vfill }}
\hbox{\vrule width 1dd
      \vbox{\hrule height 1dd
            \hbox{\vrule
                  \hbox to 70 pt{
                  \hfill\tbntry{$n$}\hfill }
                  \vrule width 1dd
                  \hbox to 30 pt{
                  \hfill\tbntry{$0$}\hfill }
                  \vrule
                  \hbox to 30 pt{
                  \hfill\tbntry{$1$}\hfill }
                  \vrule
                  \hbox to 30 pt{
                  \hfill\tbntry{$2$}\hfill }
                  \vrule
                  \hbox to 30 pt{
                  \hfill\tbntry{$3$}\hfill }
                  \vrule
                  \hbox to 30 pt{
                  \hfill\tbntry{$4$}\hfill }
                  \vrule
                  \hbox to 30 pt{
                  \hfill\tbntry{$5$}\hfill }
                  \vrule
                  \hbox to 30 pt{
                  \hfill\tbntry{$6$}\hfill }
                  \vrule
                  \hbox to 30 pt{
                  \hfill\tbntry{$7$}\hfill }
                  \vrule width 1dd
                 }
            \hrule
            \hbox{\vrule
                  \hbox to 70 pt{
                  \hfill\tbntry{$KO (\S^n)$}\hfill }
                  \vrule width 1dd
                  \hbox to 30 pt{
                  \hfill\tbntry{$\Z$}\hfill }
                  \vrule
                  \hbox to 30 pt{
                  \hfill\tbntry{$\Z_2$}\hfill }
                  \vrule
                  \hbox to 30 pt{
                  \hfill\tbntry{$\Z_2$}\hfill }
                  \vrule
                  \hbox to 30 pt{
                  \hfill\tbntry{$0$}\hfill }
                  \vrule
                  \hbox to 30 pt{
                  \hfill\tbntry{$\Z$}\hfill }
                  \vrule
                  \hbox to 30 pt{
                  \hfill\tbntry{$0$}\hfill }
                  \vrule
                  \hbox to 30 pt{
                  \hfill\tbntry{$0$}\hfill }
                  \vrule
                  \hbox to 30 pt{
                  \hfill\tbntry{$0$}\hfill }
                  \vrule width 1dd
                 }
            \hrule height 1dd
         }
     }
$$
\vskip 5 pt

Now we turn to the classification of D-brane charges
that follow from \krfin\ for various values of $p$.
The $p=9$ orientifold is nothing but Type I unoriented string theory.
Apart from the familiar D-strings, 5-branes and 32 nine-branes,
the spectrum contains $SO(32)$ D-particle discovered by Sen
\refs{\senzero, \sentwo}.
The other non-BPS states with $\Z_2$-valued charges --
a gauge instanton, a seven-brane and an eight-brane --
were found in \wk\ by means of the systematic approach via $K$-theory.
Clearly, all these results are in accordance with the formula \krfin.

The formula \krfin\ allowes us to classify stable D-brane
charges localized on the $\CO^{-} 5$-plane. Due to the Bott
periodicity, the spectrum looks very much like in Type I string
theory:
$$
\Z, \qquad {\rm D-string;}
$$
\eqn\ominus{ \Z_2, \qquad {\rm gauge\ soliton;}}
$$
\Z_2, \qquad {\rm gauge\ instanton.}
$$

Among Type IIA orientifolds, a 4-plane has the form \omorr.
It was proposed in \wk, that Type IIA D-brane charges take values
in $K(X \times \S^1) \cong K^{\pm 1} (X)$. Because of the mod 2
periodicity, the uncertainty in the degree of suspension
does not affect the answer in the complex $K$-theory. However, one
has to be more accurate in the real category. We claim (and
argue in the following sections) that the correct shift is given by
one extra suspension, {\it i.e.} in the real case Type IIA
D-brane charges are measured by the group:
\eqn\kriia{KR (\R^{9-p, p-d} \times \S^1) \cong KO(\S^{2p-d}) }
Thus, under the T-duality transformation
($p \to p-1$) the dimensions of all the $d$-branes are
reduced by one, compared to Type IIB orientifolds. It means that
the only stable objects localized on a 4-plane are D-particles
and D-instantons with charges $\Z$ and $\Z_2$ respectively.


\subsec{$\tau^2 = -1$: Symplectic Bundles and Periodicity}

So far we considered 9-branes quantized with $SO$ Chan-Paton
factors according to the choice $\tau^2 = 1$ of orientifold
projection, $\Omega^2 = 1$ in the notations of \gp. Gimon and
Polchinski explained that in Type I sting theory $\Omega^2$
acts as $(-1)$ on the 5 -- 9 strings. Hence $\Omega^2 \vert 5
\rangle = - \vert 5 \rangle$, and 5-branes must be
quantized with $Sp$ Chan-Paton factors. On the other hand,
T-dualizing four directions one would get an orientifold 5-plane
with 5-branes and 9-branes interchanged because T-duality along
the $x^i$ direction
maps $\Omega$ to $\Omega \cdot \CI_{x^i}$, and vice versa.
This implies the existence of two kinds of orientifolds $\CO^{\pm}$
with the same geometry \omorr, but different gauge groups.
Explanation of all these phenomena in terms of $K$-theory
will be the goal of the present section. As a byproduct,
we find new non-BPS 3-branes and 4-branes localized on
an $\CO^{+}5$-plane.

As we have already announced, the two choices of projection
$\tau^2 = \pm 1$ give rise to $KR$ and $KH$ groups respectively.
While the first choice was the subject of the previous subsection,
now we focus on the properties of $KH(X)$. First of all, if
the involution acts trivially on $X$, i.e. $X=X_R$, then
$KH(X_R) \cong KSp(X_R)$. This is a symplectic analog of the
relation \krtoko\ in the real case. It follows that
the $KH$-theory inherits
many properties of the $KSp$-theory. Namely, multiplication
by the generator of $KH^{-4} (\pt) \cong KSp^{-4} (\pt) = \Z$
induces periodicity isomorphisms:
\eqn\period{KH^{-4} (X) \cong KR (X), \quad
KR^{-4} (X) \cong KH (X) }
Using these formulas, one can always reduce calculation of
$KH$-groups to the real $K$-theory.

Now we return to the orientifolds \omorr\
with $\tau^2=-1$, and study the spectrum of $d$-brane
charges measured by $KH(\R^{9-p, p-d})$ with a compact
support. Using the periodicity \period, it is convenient
to rewrite \krtoko\
and \krisom\ for the symplectic case at hand:
$$
KH (X_R) \cong KSp(X_R)
$$
\eqn\khisom{ KH (X) \cong KH^{-8} (X) }
$$
KH^{p,q} (X) \cong KH^{p+1,q+1} (X) \cong KH^{p-q} (X)
$$

If $X=\pt$, the case relevant to orientifold applications,
these isomorphisms might be derived independently repeating
arguments in \atiyah\ for $\tau^2 = -1$ or via the relation
to Clifford algebras \refs{\abs, \kartwo}.

Calculation of the groups $KH(\R^{9-p,p-d})$ is similar
to the corresponding computation in the real $K$-theory.
The periodicity isomorphism (the last line in \khisom) yields
$KH(\R^{0, 2p-d-1})$ which is isomorphic to
$KSp (\S^{2p-d-1})$ in the theory with compact support.
Finally, using the standard periodicity
theorem $KSp (\S^n) = KO (\S^{n+4})$, we obtain:
\eqn\khfin{KH(\R^{9-p, p-d}) \cong KO(\S^{2p-d+3}) }
Of course, this result was expected from the consecutive
application of \period\ and \krfin.

Now we shall discuss the interpretation of the $d$-brane
charges given by \khfin.
For instance, if $p=5$, we get the following $d$-branes
localized on an $\CO^{+}5$-plane with charges:
$$
\Z, \qquad {\rm 5-brane;}
$$
\eqn\ofive{ \Z_2, \qquad {\rm 4-brane;}}
$$
\Z_2, \qquad {\rm 3-brane.}
$$
It is instructive to see how the states \ofive\ with $d<5$
can arise from the gauge bundles on the five-branes placed
at the singularity. Choosing $\tau^2 = -1$, we start with
$KH(\R^{5-d} \times \R^{4,0})$ in a ten-dimensional space-time.
Because of eqs. \khisom\ and \period, this group is isomorphic
to $KO(\R^{5-d})$ which implies orthogonal gauge bundles on 5-branes.
Indeed, $KO(\R^{5-d})$ with compact support is equivalent
to the stable homotopy group $\pi_{4-d} (O(N))$ for sufficiently
large $N$. To exhibit
this, one needs to compactify $\R^{5-d}$ by a point `at
infinity' and to regard $\S^{5-d}$ as a union of two
hemispheres intersecting over the `equator' $\S^{4-d}$.
A transition function on $\S^{4-d}$ describes $O(N)$
vector bundles over $\S^{5-d}$, hence the isomorphism
$KO(\R^{5-d}) \cong \pi_{4-d} (O(N))$. Because $\pi_0
(O(N)) = \pi_1 (O(N)) = \Z_2$ we again come to the 3-brane
and 4-brane with $\Z_2$ charges \ofive. Similar argument
can be used to demonstrate that five-branes at the
$\tau^2 = +1$ orientifold discussed earlier carry
symplectic gauge bundles, in agreement with Gimon and
Polchinski \gp. In that case, non-trivial homotopy groups
$\pi_4 (Sp) = \pi_5 (Sp) = \Z_2$ account for the $Sp$
gauge soliton and instanton \ominus.

It is important to stress here that the orientifold symmetry group
$\{1, \Omega \CI\}$ consists just of two elements.
If we rather considered a larger symmetry group,
the charges of D-branes would be classified by
another equivariant $K$-theory. For example, dividing by
the group of four elements $\{1, \CI, \Omega, \CI
\Omega \}$, one obtains a theory equivalent to
K3 compactification of Type I theory \gp. D-brane charges in the
latter theory take values in the group $KO_{\Z_2} (X)$
rather than $KR (X)$.


\subsec{Stringy Construction}

To conclude this section, we comment on the stringy
construction of new non-BPS objects. Non-supersymmetric
states \ominus\ and \ofive\ localized on orientifold
5-planes $\CO^{\mp}$ will be our main examples.

Following \wk, it is natural to propose that a $d$-brane for $d$ odd
is a bound state of a Type IIB $d$-brane and an anti-brane
exchanged by the $\Omega$ action, {\it i.e.}
$d$ could be either $-1$, 3 or 7. If nine-branes are quantized with
orthogonal Chan-Paton factors, it turns out that
the tachyon is removed by $\Omega$ projection only
for $d=-1$, 7 \wk. On the other hand, in the case
$\tau^2 = -1$, only 3 --- ${\bar 3}$ system is stable.
This is indeed what we found in \ominus\ and \ofive.

When interpreting a $d$-brane with $d$ even, one encounters
the same problem as in \wk. Namely, Neveu-Schwarz and
Ramond sectors of open $d$ -- $p$ string produce odd numbers
of fermion zero-modes. Consistent quantization of the
corresponding Clifford algebras is obstructed by
the absence of the operator $(-1)^F$ that would anti-commute
with fermionic modes. To resolve the difficulty, Witten
proposed to introduce one extra fermion zero mode $\eta$,
anti-commuting with the other fermions $w_i$. Then
the operator $(-1)^F$ can be defined as the product
$\eta \prod_i w_i$. The appearance of the zero mode $\eta$
has several effects on string dynamics. Firstly, in effect
there is no GSO projection on the string ground state because
we have enlarged the original Fock
space \refs{\senfour, \wk}. Secondly, the world-sheet path
integral has an extra factor $\sqrt2$ from the $\eta$
mode in the NS sector, so that the masses of all such
$d$-branes are $\sqrt2$ times greater than the masses of
the corresponding Type IIA D-branes. Furthermore, after
adding $\eta$ field and making the GSO projection,
we obtain chiral spinors of $SO(1,d+1)$ in the Ramond
sector of $d$ -- $p$ string. These fermions must be real or
pseudoreal to agree with the orientifold projection.
It is easy to see that this is indeed the case \cliff.
For example, $\Cl_{1,5} = \IH (4)$
confirms the existence of D-particle on the
$\CO^{-}5$-plane, in accordance with $KSp(\S^5) = \Z_2$.
In turn, an orientifold 5-plane supplemented with
an orthogonal gauge group has a $\Z_2$ charge of non-BPS
4-brane \ofive. This is in perfect agreement with
the corresponding Clifford algebra $\Cl_{1,1} = \IR (2)$ which
is real.

Relation between fermion zero modes in the Ramond
sector and Clifford algebras seems to be more profound,
and begs for further investigation.


\newsec{Dynamics of Type I D-branes}

Unlike the usual D-branes, new non-supersymmetric branes
with $\Z_2$ charges found above do not couple to massless
Ramond-Ramond fields. Of particular interest is the
question about the interaction of such states in Type I string
theory. The interaction amplitudes of Type I D-particle
can be found using the set of rules in \senfour.
Another (topological) sort of interaction could be
the discrete electric-magnetic duality in $p$ -- $q$
brane systems with $p + q =7$, as proposed by Witten \wk.

To justify the conjecture of \wk, in this section we
demonstrate the $(-1)$ monodromy in two Aharonov-Bohm
experiments:

{\it (a)} when we parallel transport a D-particle around
a 7-brane;

{\it (b)} when we parallel transport a gauge instanton
across an 8-brane.

We expect the interaction to be mediated by $p$ -- $q$ strings
and to be topological in the sense that it should not depend on small
perturbations, but must feel the relative orientation of the
brane system. The last effect can be felt only by fermions that
become massless when the branes come close to each other.
In the Neveu-Schwarz sector, the $p$ -- $q$ string zero point
energy equals $- \hf + ({\rm DN} + {\rm ND})/8 > 0$ \tasi.
Therefore, we have to focus on the fermions in the Ramond
sector where the ground state energy is always zero.

Below we study the fermions in the Ramond sector of $p$ -- $q$
string by two different methods. First, we present `stringy'
approach where the monodromy appears as a Berry's phase, and
0 -- 7 system is the most convenient example to use.
On the other hand, case {\it (b)} is the main
example of the second approach via gauge bundles.


\subsec{0 -- 7 Strings and Berry's Phase}

In order to observe the Berry's phase in the 0 -- 7 system,
we establish the degeneracy of fermion energies in the
Ramond sector when the branes coincide. Then we show
that the degeneracy is lifted once the D-particle moves away from
the 7-brane. We place the 7-brane at $x^8 = x^9 =0$ and choose
the position of the D-particle to be $x^{\m} = (0, \ldots 0,
\vec a)$, $\m >0$, where $\vec a$ is the position vector
in the $8-9$ plane. For the time being we put $\vec a = 0$.

Type I seven-brane is a bound state of a Type IIB 7-brane and
an anti-7-brane where the tachyon is projected out by $\Omega$
\wk. Therefore, Type I 0 -- 7 string spectrum contains two
copies of modes, corresponding to a 0 -- 7 string and a 0 -- ${\bar 7}$
string in Type IIB theory. Because these are oriented
strings, the fermions are complex. In what follows we will
count real fermions, {\it i.e.} we will distinguish between
0-7 strings and 7-0 strings, the fermions of the last two
being real. In total we obtain 0 -- 7, 7 -- 0, 0 -- ${\bar 7}$ and
${\bar 7}$ -- 0 strings. The world-sheet orientation reversal
$\Omega$ maps Type IIB 7-branes to ${\bar 7}$-branes,
and vice versa. Therefore, only two sets of the modes
listed above are independent: $\Omega$ identifies 0 -- 7
with ${\bar 7}$ -- 0 strings, and 0 -- ${\bar 7}$ with 7 -- 0.
Let us consider 0 -- 7 and 0 -- ${\bar 7}$ independent string sectors.

Taking into account the extra fermion field $\eta$, there
are four fermion zero modes in the Ramond sector of
the 0 -- 7 sting\foot{Discussion of the 0 -- ${\bar 7}$ sector
requires only minor modifications which we will make later.}:
$w^0$, $w^8$, $w^9$ and $\eta$. Fixing
the light-cone gauge in the $8-9$ directions,
we end up with two real fermions \gglcone.
It is convenient to combine them into the creation
and annihilation operators $d^{\pm} = \hf (w^0 \pm \eta)$
which generate two Ramond ground states \tasi:
\eqn\grounds{ \vert + \hf \rangle
\quad {\rm and} \quad \vert - \hf \rangle}

These eigenstates represent two irreducible
representations of the two-dimensional rotation
symmetry group $SO(2)$ with eigenvalues
$s = \pm \hf$ respectively. The GSO projection
keeps only one of them, the one with even fermion number.
Assuming $d^{-} \vert - \hf \rangle =0$, we end up with
the only fermion zero mode $\vert - \hf \rangle$
in the Ramond sector of the 0 -- 7 string.
The discussion of the 0 -- ${\bar 7}$ sector is very similar,
and we still get two fermion zero modes \grounds.
But this time, since 7 -- 7 and 7 -- ${\bar 7}$ vertex operators undergo
the opposite GSO projections, consistent OPE of 0 -- 7 -- ${\bar 7}$ string
triangle requires the GSO projection in the 0 -- ${\bar 7}$ sector
to be opposite to that in the 0 -- 7 sector \bgs.
Hence now we end up with the zero mode of
opposite chirality, $\vert + \hf \rangle$.
To summarize our results, in the system of coinciding
0-brane and 7-brane we have found two fermion zero modes
with quantum numbers as in \grounds.

Now we argue that the two-fold degeneracy found above
is lifted if we perturb the system by small displacement
of the D-particle, $\vec a \ne 0$. Because prior to
the gauge fixing fermion zero modes
$w^0$, $w^8$, $w^9$ and $\eta$ were in the
same representation of the four-dimensional Clifford
algebra $\Cl_{1,3}$, we can choose the $SO(2)$ symmetry
group in the previous paragraph to be the rotation symmetry
in the 8 -- 9 plane. Furthermore, physical states \grounds\
must satisfy the super-Virasoro constraint:
\eqn\svirasoro{G_0 \vert \psi \rangle = 0}
which, on the ground states, reduces to the two-dimensional
Dirac equation $p_{\m} w^{\m} \vert \psi \rangle
\simeq \D \psi=0$. Because the states \grounds\
have opposite $SO(2)$ chirality, they have different
eigenvalues. It means that degeneracy is lifted as long as
$\vec a \ne 0$, {\it i.e.} when 0 -- 7 string has finite
length.

After all, we have the two-level system with parameter
space $\{ \vec a \}$, such that levels cross
\foot{We assume that perturbation of energy levels
is first order in $\vec a$. Direct calculation in
the end of this subsection will confirm this assumption.}
at the single point $\vec a =0$.
This is sufficient information to deduce
the Berry's phase acquired by the ground state during
adiabatic transport of $\vec a$ around the origin \berry.
To the first order in perturbation, the general
Hamiltonian describing the two levels \grounds\
in the real representation of $SO(2)$ can be expressed
in terms of real Pauli matrices:
\eqn\hamilt{H (\vec a) =  \hf \pmatrix{a^8 & a^9 \cr
a^9 & - a^8 \cr} =  \hf \vec \sigma \cdot \vec a}
Note, the same Hamiltonian describes 3d spin with $S=\hf$
in the external magnetic field $(a^8, 0, a^9)$, and the
so-called dynamical Jahn-Teller effect. It is important
to stress here that because of the reality condition
``the Berry's phase'' is actually a discrete number
($0$ or $\pi$) rather than a continuous phase.
And the eigenfunction of the pure state $\vert s \rangle$
can change the sign via mixing with the orthogonal state
during the adiabatic transport, {\it e.g.}:
$$
\vert + \hf~ (\theta) \rangle =
\cos ({\theta \over 2}) \vert + \hf \rangle +
\sin ({\theta \over 2}) \vert - \hf \rangle
$$
This is an eigenfunction of the Hamiltonian \hamilt\ where
$\vec a = (a \cos{\theta}, a \sin{\theta})$.
Analogous pattern takes place in the dynamical Jahn-Teller
effect. The topological phase is given by half the `solid
angle' that the adiabatic path subtends at the degeneracy point,
{\it i.e.} $\varphi = \hf (2 \pi) = \pi$. This leads to the
expected monodromy $\exp(i \varphi) = -1$.

In order to see how the Hamiltonian \hamilt\
follows from string dynamics, it is
convenient to consider string coordinates ($i=\{8,9\}$):
$$
X^{\m} (z, \bar z) = X^{\m} (z) + X^{\m} (\bar z) =
-i {a^{\m} \over 2 \pi} \ln({z \over \bar z}) +
{\rm oscillators}
$$
in the T-dual picture \tasi:
$$
\tilde X^{\m} (z, \bar z) = X^{\m} (z) - X^{\m} (\bar z) =
-i \a' p^{\m} \ln(z \bar z) + {\rm oscillators}
$$
where $p^{\m} = a^{\m} / (2 \pi \a')$. Therefore,
small perturbation of the `Dirac equation' \svirasoro\
leads to the effective Hamiltonian \hamilt\
in the representation $w^8 = \sigma_x$, $w^9 = \sigma_z$.
It follows that energy gap between two states \grounds\
is proportional to $a$ which confirms our assumption
about conical crossing of energy levels at the origin.


\subsec{Approach via Gauge Theory}

Now we turn to another face of the $p$ -- $q$ strings where
the branes are represented by topological defects in the
gauge bundles on 9-branes. This approach is reminiscent
of the $K$-theory construction \tachyon.
Since tadpole cancellation requires 32 nine-branes to present
in Type I string theory from the very beginning \gp, we don't
need to invoke extra anti-branes to construct the $p$ -- $q$ system.
Following this reasoning, we study $\CN =1$ effective $SO(32)$ gauge
theory on the world-volume of parallel 9-branes:
\eqn\action{\Tr \int (\hf F_{\mu \nu}  F^{\mu \nu}
+ i \bar \Psi [ \D, \Psi])} 
where $\Psi$ is the Weyl fermion, and $F_{\m \n}$ is the
field strength of the gauge field. In general, the
background of $p$- and $q$-branes system ($q=7-p$) is
given by:
\eqn\backgra{A_{\mu} = \pmatrix{A_{\mu}^{(p)} & 0 \cr
0 &  A_{\mu}^{(q)} \cr}}
and vanishing fermion field. 
The gauge connection $A_{\mu}^{(p)}$ describing the $p$-brane
depends on $(9-p)$ coordinates $x^i$ transverse to the $p$-brane.
This is in accordance with the fact that the corresponding
bundle $E_{(p)}$ (together with the trivial bundle of rank 0)
represents the non-trivial element of $KO(\R^{9-p})$.

In this language, the fermions in the Ramond sector of
$p$ -- $q$ strings are represented by the off-diagonal
blocks $\psi$ and $\psi^{\dag}$ of the fermion field \holiwu:
\eqn\backgrf{\pmatrix{0 & \psi \cr \psi^{\dag} & 0 \cr}}
The Weyl fermion $\psi$ is a section of $E_{(p)} \otimes E^{\ast}_{(q)}$.

Now it is convenient to focus on the $p=8$ ($q=-1$) system.
We are interested in the zero modes of $\psi$ when the gauge
instanton and the 8-brane are placed at the same point $x^9=0$.
An advantage of 8-branes is that rank of the bundle $E_{(8)}$
is equal to 1, {\it i.e.} we don't have to worry about the
corresponding indexes. Hence, according to the index theorem
\refs{\indexone, \indextwo}, in the sector with non-trivial instanton
numbers, the Dirac operator $\D (A_{(-1)}) \oplus \D (A_{(8)})$
has one zero mode of definite chirality with respect to the
operators $( \prod_{\mu=0}^9 \Gamma^{\mu} )$ and $\Gamma^9$.
Here it is important that we deal with orthogonal gauge group.
Consider perturbation of this system by small displacement of
the 8-brane: $x^9 \to x^9 - a$. Effective action for the zero
mode $\psi_0$ follows from \action:
$$
\int \psi_0^{\dag} (\Gamma^9 a) \psi_0
$$
Because $\psi_0$ satisfies $\Gamma^9 \psi_0 = + \psi_0$,
the eigenvalue of the Dirac operator
$$
\D_a = \D (A_{(-1)}) + \D (A_{(8)}) + \Gamma^9 a
$$
is equal to $+a$, and changes its sign as
the instanton crosses the 8-brane. Hence, fermion contribution to
the amplitude $(\Det \ i \D_a)^{\hf}$, defined as the product of
the half of the eigenvalues, also changes the sign. The other choice
of the disconnected component of the orthogonal group, corresponding
to the opposite sign in $\Gamma^9 \psi_1 = - \psi_1$, would result
in the fermion mode $\psi_1$ which always remains massive in the
neighborhood of $a \simeq 0$. Therefore, it would not affect
the path integral, as well as other massive modes.

Like in the approach via Berry's phase, the $(-1)$ monodromy
is produced by the fermions which become massless when the
branes coincide. Actually the two methods are equivalent
and are based on the spectral flow of the Dirac operator.

In general, using the Thom isomorphism, it is convenient to reduce
the problem to two dimensions. Then, a 7-brane and a (-1)-brane
become a gauge instanton, while a 0-brane and an 8-brane transform
into a two-dimensional soliton. The world-line of the gauge soliton
is one-dimensional curve, say $x^1=a$. We want to
demonstrate that the sign of the instanton amplitude is
reversed in crossing the curve $x^1 = a$.
Even though this system is very similar to the $(-1)$ -- 8 case
discussed above, we use a different argument to show that odd
number of eigenvalues of the Dirac operator $\D_a$ change sign.
As usual, to find the spectral flow uder deformation from
$\D_{- \infty}$ to $\D_{+ \infty}$, one has to promote $a$
to the independent coordinate, $\CD = \D_a + \Gamma^a \p_a$. Then,
the spectral flow of $\D_a$ is equal to the index of $\CD$ \atanomaly.
Now, to complete the proof, we show that ${\rm ind} (\CD)$
represents a non-trivial element in $K$-theory\foot{Here we
use equivalence of the topological and the analytical indices
\refs{\indexone, \indextwo}. }. Since $A_{(0)}$ depends only on
$(x^1 - a)$, the contribution from the $a$ `direction' is the same
(up to relative sign) as the contribution from the gauge soliton.
Therefore, we end up with ${\rm ind} (\D_{(-1)})$ corresponding to
the gauge instanton class in $KO$-theory.


\newsec{Orientifolds of type (ii):
$\R^{p+1} \times (\R^{9-p} / (-1)^{F_L} \cdot \CI_{9-p})$}

Now we consider Type IIB orientifolds where involution is
combined with the perturbative symmetry group $(-1)^{F_L}$.
Acting on 9-branes, it maps a pair of bundles $(E,F)$ to its
`negative' $(F,E)$, in the sense $(E,F) = E-F$. According
to \wk, charges of $d$-branes localized at the singularity
take values in the corresponding $K$-theory group
$K_{\pm} (\R^{9-p,p-d})$ that will be the main subject of this
section. Because calculation of $K_{\pm} (\R^{p,q})$ involves
both unreduced and reduced K-theories, notations in this
section slightly differ from the rest of the paper.
Namely, we restore the conventional notation $\tilde K(X)$
for reduced cohomology of $X$ with the base point, while
the symbol $K(X)$ will denote unreduced $K$-theory.

It has been shown by M.~J.~Hopkins that calculation of
$K_{\pm}$-groups can be carried out in terms of the usual
$\Z_2$-equivariant $K$-theory by means of the formula \wk:
\eqn\hopkins{
\tilde K_{\pm} (X) \cong K^{-1}_{\Z_2} (X \times \R^{1,0})}
Note that we always imply cohomology theory with compact support.

Since the right-hand side of \hopkins\
represents a functor in the complex
category, multiplication by the Thom space of $\IC$
(or $\IC / \Z_2$) induces the periodicity isomorphisms:
\eqn\thperiod{
\tilde K_{\pm} (\R^{p,q}) \cong \tilde K_{\pm} (\R^{p,q+2}),
\qquad \tilde K_{\pm} (\R^{p,q}) \cong \tilde K_{\pm} (\R^{p+2,q}) }
Therefore, $\tilde K_{\pm} (\R^{9-p,p-d})$ depends only on parity of
$p$ and $d$. Consider first the case when $p$ is even.
Application of the Hopkins' formula \hopkins\ leads to
the equivariant group:
$$
K_{\pm} (\R^{9-p,p-d}) \cong  K^{-1}_{\Z_2} (\R^{10-p,p-d})
$$
which, by the periodicity theorem \thperiod, gives the answer
for $d$-brane charges ($p$-even):
\eqn\kpmeven{\tilde K_{\pm} (\R^{9-p,p-d}) \cong 
K^{-1}_{\Z_2} (\R^{10-p,p-d}) \cong K^{d-1}_{\Z_2} (\pt) }
The last group is isomorphic to the representation ring
$R[\Z_2]$ if $d$ is odd, and is trivial if $d$-even
\refs{\segal, \eqlect}. However, $p$-even is not the case
relevant to Type IIB orientifolds discussed in \refs{\senzero,
\senone}.

To determine $\tilde K_{\pm} (\R^{9-p,p-d})$ for
$p$-odd, we employ the exact sequence \tri\ to the pair
$(\B^{9-p,p-d}, \S^{9-p,p-d})$:
\eqn\exactone{
\ldots \to K^n_{\Z_2} (\B^{9-p+1,0}, \S^{9-p+1,0}) \to 
K^n_{\Z_2} (\B^{9-p+1,0}) \buildrel \la \over \rightarrow 
K^n (\S^{9-p+1,0} / \Z_2) \to \ldots }
where we used the suspension isomorphism to substitute $d$ by
a $\Z_2$-graded index $n$. Let us analyze each term
in the part of the sequence \exactone. The first term
is obviously isomorphic to the $K$-theory $\tilde K_{\pm} (\R^{9-p,p-n})$
with compact support we are interested in.
Since $\B^{9-p+1,0}$ is equivariantly contractable, we
also get $K^n_{\Z_2} (\B^{9-p+1,0}) \cong K^n_{\Z_2} (\pt)$,
the second term in \exactone.
Therefore, the sequence \exactone\ relates
groups in question to the cohomology theory of the horizon
$H \cong \RP^{9-p}$ \adams:
$$
K^n (\RP^{9-p}) = \cases{\Z \oplus \Z_{2^r}, &
$r= \left[{9-p \over 2}\right]$, $n$ even; \cr
0, & $n$ odd.\cr}
$$
Careful analysis of the ring structure shows that $\la$
in \exactone\ maps
the generator of $K_{\Z_2} (\B^{9-p+1,0})$ to the generator
of $K^n (\RP^{9-p})$. Finally, it follows that $\tilde K_{\Z_2}
(\R^{9-p+1,0}) = \Z$ and $K^1_{\Z_2} (\B^{9-p+1,0}) = 0$.
It is convenient to list the results in the following table:
\vskip 5 pt
$$
\def\tbntry#1{\vbox to 23 pt{\vfill \hbox{#1}\vfill }}
\hbox{\vrule width 1dd
      \vbox{\hrule height 1dd
            \hbox{\vrule
                  \hbox to 70 pt{
                  \hfill\tbntry{$\tilde K_{\pm} (\R^{9-p,p-d})$}\hfill }
                  \vrule width 1dd
                  \hbox to 30 pt{
                  \hfill\tbntry{$d$ even}\hfill }
                  \vrule
                  \hbox to 30 pt{
                  \hfill\tbntry{$d$ odd}\hfill }
                  \vrule width 1dd
                 }
            \hrule height 1dd
            \hbox{\vrule
                  \hbox to 70 pt{
                  \hfill\tbntry{$p$ even}\hfill }
                  \vrule width 1dd
                  \hbox to 30 pt{
                  \hfill\tbntry{$0$}\hfill }
                  \vrule
                  \hbox to 30 pt{
                  \hfill\tbntry{$R[\Z_2]$}\hfill }
                  \vrule width 1dd
                 }
            \hrule
            \hbox{\vrule
                  \hbox to 70 pt{
                  \hfill\tbntry{$p$ odd}\hfill }
                  \vrule width 1dd
                  \hbox to 30 pt{
                  \hfill\tbntry{$\Z$}\hfill }
                  \vrule
                  \hbox to 30 pt{
                  \hfill\tbntry{$0$}\hfill }
                  \vrule width 1dd
                 }
            \hrule height 1dd
         }
     }
$$
\vskip 5 pt

Since only odd values of $p$ are possible in Type IIB sring
theory, $d$-brane charges localized on 
$\R^{p+1} \times (\R^{9-p} / (-1)^{F_L} \cdot \CI_{9-p})$
orientifolds are classified by the second line of the table.
Some states on such orientifolds
have already been discussed in the literature. For example,
if $p=9$, we obtain the standard spectrum of Type IIA string
theory: even-dimensional branes of arbitrary integer charge.
Notice, we obtain a direct argument that D-brane
charges in Type IIA string theory are classified by $K(\Sigma X)$,
regarding it as $(-1)^{F_L}$ orientifold of Type IIB theory.

Recently, non-BPS D-particle on such an $\CO 5$-plane has also
been discussed by Sen \refs{\senzero, \senone}.
Note, charge of the D-particle on the orientifold
$\R^6 \times (\R^4 / \Omega \cdot \CI_4)$ takes value in
$\Z_2$, while charge of the D-particle that lives on the
$\R^6 \times (\R^4 / (-1)^{F_L} \cdot \CI_4)$ orientifold
can be arbitrary integer. Actually there is no discrepancy
here, because $K$-theory classifies charges of topologically
stable objects only at weak coupling. On the contrary,
S-duality which relates the two types of orientifolds
inverts string coupling constant, {\it i.e.} maps type $(i)$
orientifold at weak coupling to type $(ii)$ orientifold
at strong coupling. Hence, spectra of states may not
be the same. Below we also show that masses of the states
differ by a factor of $\sqrt2$.


\subsec{Stringy Construction}

Using analogy with Type IIA string theory, it is not
difficult to provide string theory construction of the new states.
In the case $p=5$ this was done by Bergman and Gaberdiel
\bergman. Following the notation of \refs{\senone, \bergman}, we define
Type IIB closed string boundary state in the light-cone gauge:
$$
\vert Bd, \eta \rangle = \exp \{
\sum_{n>0} {1 \over n} [\a^I_{-n} \tilde \a^I_{-n} -
\a^i_{-n} \tilde \a^i_{-n}] + i \eta \sum_{r>0}
[\psi^I_{-r} \tilde \psi^I_{-r} - \psi^i_{-r}
\tilde \psi^i_{-r}] \} \vert Bd, \eta \rangle^{(0)}
$$
where $\eta = \pm$ and $n \in \Z$. Index $r$ labels
the fermion oscillators and runs over integers or
half-integers ($r \in \Z + \hf$)
depending on the sector: untwisted or twisted (U/T); NS or R;
Neumann ($i = 1, \ldots,d+1$) or Dirichlet ($I = d+2, \ldots ,8$)
boundary conditions. As usual, we choose NS -- NS sector
ground state $\vert Bd, \eta \rangle^{(0)}$ to be odd
under $(-1)^{F_L}$ and $(-1)^{F_R}$. Therefore, NS -- NS
boundary state for new $d$-branes must have the same
form as for ordinary Type II D-branes. On the other
hand, because $d$ is even, there are no R -- R boundary
states invariant under $(-1)^{F_L}$ in the untwisted
sector of Type IIB string. Nevertheless, the closed
string spectrum includes a twisted sector where the
left-GSO projection is opposite, and we {\it do} get
invariant R -- R boundary states for $d$-even.
It means that the even-dimensional branes found above can be
interpreted as twisted states localized at the orientifold plane.
Combining the contributions of NS -- NS and
R -- R sectors, we obtain:
$$
\vert Bd \rangle = ( \vert Ud, + \rangle_{{\rm NS-NS}} -
\vert Ud, - \rangle_{{\rm NS-NS}} ) + ( \vert Td, +
\rangle_{{\rm R-R}}
+ \vert Td, - \rangle_{{\rm R-R}} )
$$
This boundary state has precisely the same form as
the boundary state of the ordinary Type IIA $d$-brane.
Hence, masses of the corresponding $d$-branes
are also equal (there is no extra
factor $\sqrt2$). The authors of \bergman\ also
noticed that masses of D-particles on orientifolds
of type $(i)$ and type $(ii)$ are different. Here
we observe that not only the masses of all other
states do not match, but also their charges are
different. Again, this confirms the idea that we can not
simply follow from weak to strong coupling.


\newsec{Miscellany}

\subsec{Orientifolds of type (iii) and Relation to Clifford
Algebras}

In the previous sections we considered Type IIB orientifolds
where we divided either by $\Omega$ or by $(-1)^{F_L}$
perturbative symmetry group. Amalgamating the two cases
we obtain orientifolds of type $(iii)$:
\eqn\typetri{ \R^{p+1}
\times (\R^{9-p} / \Omega (-1)^{F_L} \cdot \CI_{9-p})}
Even though we will not try to develop $KR_{\pm}$-theory
of such orientifolds, a few comments are in place here.
In order to calculate groups $KR_{\pm} (X)$, we need an
analog of Hopkins' formula \hopkins\ in the real category,
something like:
\eqn\kpmguess{KR_{\pm} (X) \cong KR_{\Z_2} (X \times \R^{1,1})}
Validity of such formula would strongly depend on the definition
of the appropriate $K$-theory. For example, \kpmguess\ would be
true if we defined $KR_{\pm} (X)$ as a cohomology theory of $X$
that fits into the following exact sequence (the way similar to
how M.~J.~Hopkins defined $K_{\pm} (X)$ group):
\eqn\kpmlong{
\ldots \to KR_{\Z_2} (X) \to KR (X) \to KR_{\pm} (X) \to \ldots }
Using the five lemma for \kpmlong~ and the exact sequence
in $KR_{\Z_2}$-theory for the pair
$(X \times \R^{1,0}, X \times (\R^{1,0} - \pt ))$ we come
to \kpmguess. However, \kpmlong\ might not be the suitable
definition of $KR_{\pm}$ for orientifold applications.

There is another evidence to \kpmguess\ based on the relation
between $K$-theory of $n$-dimensional vector space $X^n$
and the corresponding Clifford algebra $\Cl_n$ \kartwo.
In fact, in the present paper we are mainly interested
in flat space-time orientifolds where $X=\R^{p,q}$.
For this reason, in the rest of this subsection we make a
short digression on the Clifford algebras of such spaces.

If we define $A_n$ to be the Grothendieck group of
graded $\Cl_n$-modules modulo those extendable to
$\Cl_{n+1}$-modules\foot{The inclusion map $\Cl_{n}
\to \Cl_{n+1}$ is induced by $X^n \to X^n \oplus \IR$.},
then there exists an isomorphism \refs{\abs,\kartwo,\askew}:
$$
A_n \cong K (X^n)
$$
We can use this isomorphism twice, first to convert
the problem to the algebraic one, and then to read off the
answer for $K (X^n)$. In general, analysis of Clifford algebras
is very simple, and many results in the previous sections
become manifest once translated to the algebraic language.
For example, let us prove the periodicity isomorphism
\krisom, namely $\Cl_{p,q} \cong \Cl_{p-4, q+4}$, $p>4$.
Take an orthonormal basis of $\IR^{p,q}$ generated by
matrices $\g_{\m}$, such that\foot{Note, here we use
the equivalence between the Clifford algebra of the real
space $\R^{p,q}$ with involution $\tau$, $\tau^2=+1$,
and the Clifford algebra of the linear space $\IR^{p,q}$
with signature $(p,q)$ \atiyahr.}:
\eqn\gammas{\g_{\m} \g_{\n} + \g_{\n} \g_{\m} = 2 g_{\m \n}}

Now, we define:
$$
\cases{ \g_{\m}' = \g_{\m} (\prod_{\n =1}^4 \g_{\n}), &
if $\m = 1 \ldots 4$; \cr
\g_{\m}' = \g_{\m} & otherwise. \cr}
$$
Then, according to \gammas, the subset $\{ \g_{\m}' \}$ of
$\Cl_{p,q}$ generates $\Cl_{p-4, q+4}$. QED.

Involutions on $X^n$ induce (anti-)automorphisms
of the corresponding Clifford algebra $\Cl_n$, and the latter
are classified \cliff. In the orientifolds \typetri\
of type $(iii)$ the involution maps a pair of gauge bundles
$(E,F)$ to $(\overline F, \overline E)$. Since the tachyon
\tachyon\ defines a scalar product on the spin bundle
$S_{+} \oplus S_{-}$, it suggests that the involution
induces reversion automorphism of the Clifford algebra $\Cl_{p,q}$.
Calculation of the corresponding automorphism groups gives
an independent argument to \kpmguess. To be specific we
mention an intriguing example of a non-BPS state:
a 3-brane with $\Z_2$ charge is localized on the 7-plane.
However we will not pursue the analysis any further.


\subsec{AdS Orbifolds}

In the second section we briefly mentioned the AdS/CFT
correspondence \malda, which
relates the conformal gauge theory on branes placed at the orbifold
singularity and supergravity on the horizon manifold $H$.
It would be interesting to investigate further implications of this
duality in terms of $K$-theoretic relation \tri\ between
$X$ and $H$, {\it cf.} \nshorizon.

Let us consider an example of $\Z_3$ AdS orbifold which
is dual to $\CN=1$ superconformal field theory.
Namely, we study Type IIB compactification on
${\rm AdS}_5 \times (\S^5 / \Z_3)$ where the Lens space
$H = L^2 (3) = \S^5 / \Z_3$ is a genuine horizon in the
sense of \refs{\malda, \nshorizon}.
It is dual to $SU(N)^3$ gauge theory on the
boundary ( = the gauge theory on $N$ parallel 3-branes placed
at the orbifold singularity) with nine chiral multiplets
in the bifundamental representation of the gauge factors
\refs{\dgmr, \lnv}. This SCFT has discrete global symmetry
group \grw:
\eqn\ztrich{(\Z_3 \times \Z_3) > \!\!\! \triangleleft \Z_3}
where $\Z_3$ factors are generated by $A$, $C$ and $B$ such that:
$$
A^{-1} B^{-1} AB=C
$$

Extended objects in the boundary theory which are charged under
the discrete symmetry group \ztrich\ can be understood as Type IIB
branes wrapped on various cycles in $H = S^5 / \Z_3$.
Because the horizon $H$ has non-trivial homology groups
$H_1(H) = H_3(H) = \Z_3$, we end up with
even-dimensional objects propagating in ${\rm AdS}_5$ with
charges given by \ztrich. Let us focus, say, on
membranes which look like gauge strings on the boundary. There
are three types of membranes corresponding to different $\Z_3$
factors in \ztrich\ --- one can make a membrane
by wrapping a 3-brane on a 1-cycle in $H$, and by wrapping
a D5-brane or a NS5-brane on 3-cycles respectively.
The charge of the NS5-brane corresponds to the last factor
in \ztrich, and accurate analysis shows that it does not
commute with the other D-brane charges. Since in the
present paper we deal with ordinary topological $K$-theory
which does not take into account the Neveu-Schwarz $B$-field
\foot{Note, in our discussion $K^{\ast} (X)$ is always
a commutative ring.},
we don't expect to see the last $\Z_3$ charge factor
in \ztrich. Indeed,
calculation of the $K$-group of the Lens space $H$ gives
\klens:
$$
K(H) = (\Z_3)^2 \cong H^{{\rm even}} (H, \Z)
$$
Complete agreement with the ordinary cohomology theory tells us that
$K$-theory does not supply new objects for this orbifold example.


\subsec{Toric Varieties}

In fact, the result of the previous subsection is not very surprising.
A number of space-time manifolds $X$
(including the models of \nshorizon)
are birationally equivalent to smooth toric varieties. Vector
bundles over such $X$ have simple combinatorial description
on the dual lattice (see {\it e.g.} \sharpe), and $K(X)$ can be
examined in the same way \tor. By Lemma 1 of \tor, $K(X)$
is free of torsion, that is the Chern character map:
\eqn\chern{ch \colon K (X) \to H^{{\rm even}} (X, \Z)}
is an isomorphism \ah. Restriction of bundles to hypersurfaces
and complete intersections in toric varieties enlarges the range
of possible applications. More generally, \chern\ holds for
CW complexes of low dimension \karone.


\newsec{Summary}

As we have seen, $K$-theory is a powerful tool which
helped us to study charges of non-BPS D-branes
localized on the following types of orientifolds:

$(i)$ For the orientifolds of the form \omor, two choices of the
projection ($\tau^2 = \pm 1$) lead to different $K$-theories:
$KR (X)$ and $KH(X)$ respectively. In the case of flat space-time
orientifolds \omorr, we calculated these groups with the result
\krfin, \khfin.
For example, we found new D-brane charges \ominus\ and \ofive\
localized on orientifold 5-planes $\CO^{-}$ and $\CO^{+}$.
String theory construction of the new states with $\Z_2$ charges
was also discussed. In general, odd-dimensional $d$-branes
are represented by $d$ -- ${\bar d}$ configurations in
Type IIB theory, while the description of $d$-branes with
$d$-even involves extra fermion zero mode $\eta$.
It would be interesting to further investigate the dynamics
of such states either by topological methods of section 4,
where we proved the discrete electric-magnetic duality in Type I
theory \wk, or via direct computation of string amplitudes \senfour.

$(ii)$ In this case, calculation of the groups
$K_{\pm} (\R^{9-p,p-d})$, $p$-odd, resulted in the
spectrum of even-dimensional $d$-branes with arbitrary
integer charges, like in Type IIA theory. These states are
simply twisted states localized on $(-1)^{F_L}$ orientifolds.

$(iii)$ Our discussion of $\Omega \cdot (-1)^{F_L}$ orientifolds
is by no means complete. In order to calculate $KR_{\pm} (X)$,
we conjectured the isomorphism \kpmguess\ and made some
arguments in favor of it. For the seven-plane example,
it predicts the existence of 3-branes with $\Z_2$-valued charge.

Finally, we argued that $K$-theory of smooth (toric)
compactifications and their orbifold limits
does not supply new objects.

One can generalize the present analysis
to other $\CM$, say tori. Another aspect, which is not quite
clear yet, is the relation to Clifford algebras mentioned
in sections 3.3 and 6.1.


\vskip 30pt
\centerline{\bf Acknowledgments}

I am very grateful to C.~Bachas, W.~Browder, M.~J.~Hopkins,
I.~R.~Klebanov, S.~Martin, A.~Schwarz and especially to E.~Witten
for interesting and illuminating discussions/correspondence.
It is pleasure to thank Harvard University for financial support
and hospitality while the manuscript was being completed.
The work was supported in part by grant RFBR No 98-02-16575 and
Russian President's grant No 96-15-96939.

\listrefs
\end